# Attention-Enhanced U-Net for Accurate Segmentation of COVID-19 Infected Lung Regions in CT Scans

Amal Lahchim[1], Lazar Davic[2]

[1] School of Engineering, University of Kragujevac

[2] kragujevac, Serbia.

## Abstract

In this study, the focus is on developing a robust methodology for automatic segmentation of infected lung regions in COVID-19 CT scans utilizing advanced CNNs. The proposed model is based on a modified U-Net architecture with attention mechanisms, data augmentation, and postprocessing techniques, achieving high segmentation accuracy and boundary precision. The dataset was sourced from publicly available repositories, processed, and augmented to increase its diversity and generalizability. The approach was evaluated quantitatively, resulting in a Dice coefficient of 0.8658 and mean IoU of 0.8316. The proposed model is compared to existing methods through comparative analysis, clearly demonstrating its superiority in handling data variability and achieving precise segmentation. Future work involves expanding the dataset, applying an alternative 3D segmentation approach, and improving the model for clinical deployment.

*Keywords:* **COVID-19, U-Net architecture, CNNs, attention mechanisms, data augmentation**

## Introduction

As a result of the current global COVID-19 pandemic, it has placed unprecedented pressure on global healthcare systems, leading to the fast-tracking of diagnostic and therapeutic technologies (1). Medical imaging, especially CT, has emerged as an important tool for understanding and controlling the disease among these (2). In addition to helping to diagnose COVID-19, CT imaging also provides information on how severe the infection is and how widely the lungs are affected because that information is relevant to planning appropriate treatment strategies (3). CT scans are supportive for COVID-19 diagnosis and are essential in assessing the severity of the disease in patients (4). Models that can detect COVID-19-related findings and infer the pattern of infection may help enhance the diagnosis and treatment, particularly in regions where there is a deficit of expert radiologists (5). In particular, automatic segmentation of infected lung regions from CT scans is important for the localization and quantification of affected areas in order to objectively assess disease progression and monitor disease progression(6). In this report, we describe a methodology for the automatic segmentation of infected lung regions based on convolutional neural networks (CNNs) using a network architecture that is a variation of the U-Net, augmented data, and post-processing that results in high accuracy, efficiency, and reproducibility for segmentation tasks.

## Dataset and Data Preprocessing

**1. Description of the Task:**

The goal is to create a solid and rapid segmentation technique for infected regions in lung CT slices. The segmentation process produces clear and accurate maps of infection-affected lung regions to support quantitative analysis for clinical and research purposes, as well as scalability for real-world clinical deployment.

**2. Dataset Description:**

The data used in this study was publicly available on COVID-19 CT repositories hosted on Coronacases.org, Radiopaedia.org, and the Zenodo Repository. It consists of five key components: 'ct_scans,' 'metadata.csv,' 'infection_mask,' 'lung_and_infection_mask,' and 'lung_mask.' In this study specifically, the focus is on the CT scans and their associated 'infection_mask,' i.e., the portions of the CT that are infected with COVID-19. The dataset consists of 20 CT scans with dimensions of $512 \times 512 \times 301$ (height, width, and number of slices). There are also expert-provided masks that outline the lungs and infected regions (7)(8), and the images are in 12-bit grayscale and are available in DICOM and NIfTI formats.

### 3. Data Visualization:

Significant variability among the CT scans was evaluated in the form of a histogram to define the distribution of Hounsfield Units (HU) present in the scans. The HU values vary from -1000 (air) to about 1500 (bone and other densely packed structures), with lung parenchyma and soft tissues producing peak values. The way this visualization is used drastically affects preprocessing decisions. It helps in choosing normalization techniques to ensure the data appropriately represents anatomical features, that is, to identify intensity ranges. For instance, min-max normalization can produce a pixel value scale between [0, 1], and Z-score normalization is the process of scaling a pixel value relative to its mean and standard deviation. In addition, HU thresholding can be applied to narrow the range of interest, such as [-1000, 1500], instead of using the entire HU range. As seen in Figure 1, this prevents noise and unwanted data.

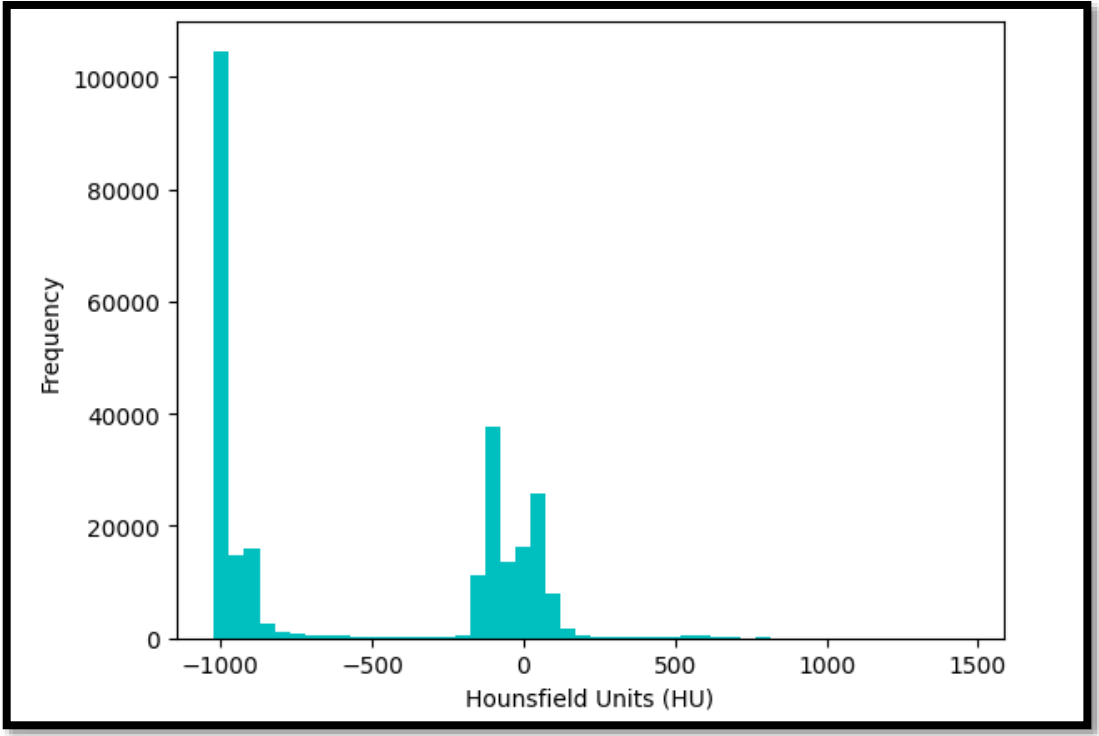

**Fig1. Frequency Distribution of Hounsfield Units (HU) in CT Scans**

**4. Data Preprocessing:**

A step that is critical to getting inputs ready for the model is preprocessing. Pixel intensities were normalized to [0, 1] first to make the values numerically stable during training. To reduce computational complexity while preserving the relevant anatomical details, images were resized to 128 × 128 pixels. The annotations were binarized so that any annotation values could only be 0, representing the background, and 1 for infected regions. Standard resolution was also spatially resampled across different scans. The preprocessing ensured that the dataset was consistent and ready to train a convolutional neural network in this case (see Figure 2).

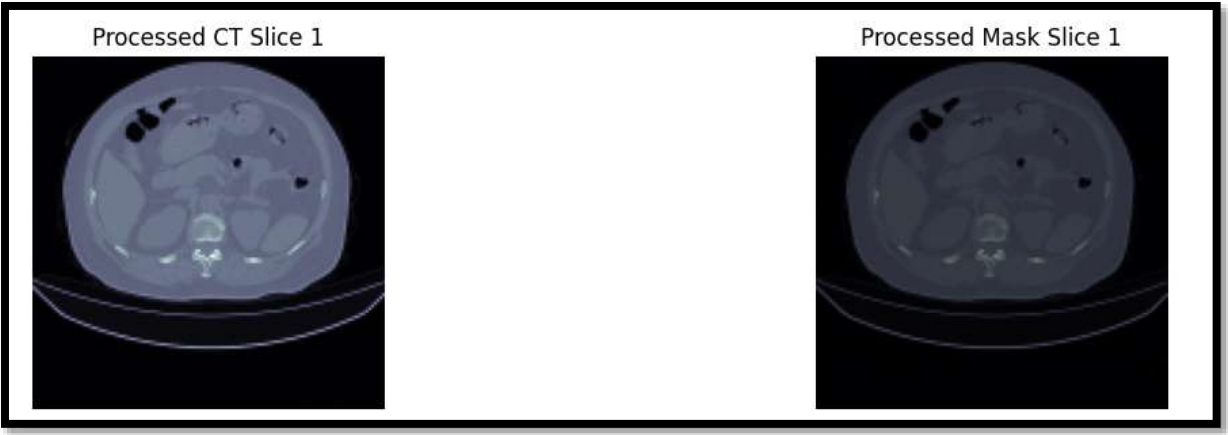

**Fig2. Visualization of Processed CT Slice and Corresponding Mask**

**Augmentation and Post-Processing:**

**1. Augmentation and Pipeline:**

A safe augmentation pipeline was implemented to increase dataset diversity and improve model generalization. Rotations were restricted to ±5 degrees, along with minor translations and scalings, as these translate into anatomical variations. Elastic transformations with reduced intensity, Gaussian blurring for smoothness, and brightness and contrast changes were applied to mimic varying imaging conditions. After augmentation, 2252 CT slices and masks were obtained, resized to 128 × 128 pixels with one channel.

The segmentation model was trained on both the original dataset and the augmented dataset to analyze the impact of data augmentation. Figure 2 demonstrates how much augmentation improved the results.

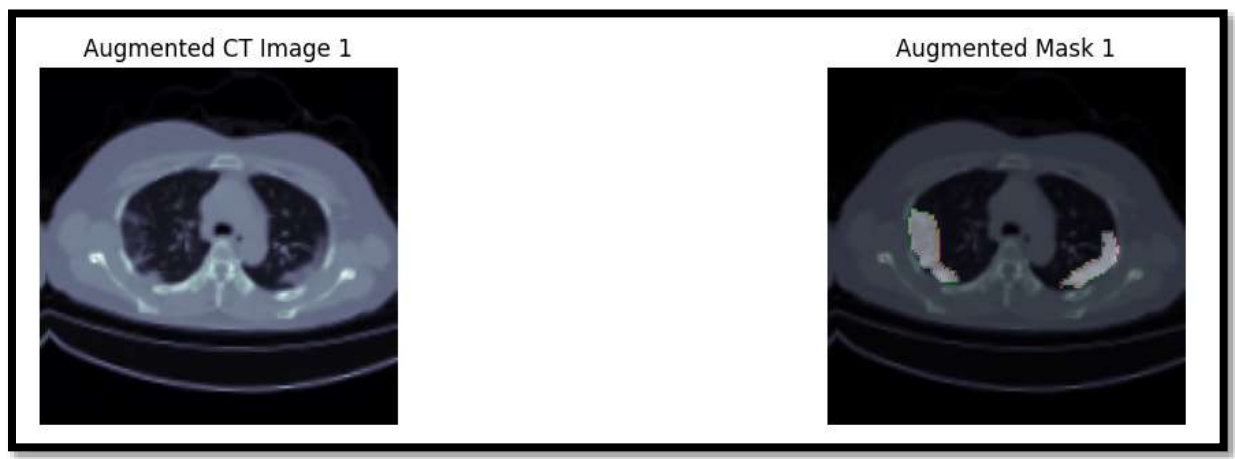

**Fig3. Visualization of Augmented CT Image and Corresponding Mask**

**2. Post-Processing:**

The predicted segmentation masks were post-processed to refine the prediction to a segmented image of this object, as well as to ensure accurate boundary delineation. Several critical steps were necessary for the post-processing pipeline. Next, we binarized the predicted masks using a threshold (e.g., 0.3) such that pixel values greater than the selected threshold were marked as part of the infected areas. Finally, morphological operations using a 3x3 kernel of opening and closing were applied to remove small noise and to fill small holes in the masks. To get rid of small objects smaller than a little size threshold (e.g., 10 pixels), connected component analysis was applied to keep only the supposed significant regions. Finally, the processed masks were resized to the original dimensions of the input CT slices. With this post-processing pipeline, the quality of the predicted masks is significantly improved, and artifacts are eliminated; segmentation accuracy is also improved.

## Tools and Methods:

The project employed a comprehensive set of techniques and tools to address the task of infected lung region segmentation. Python served as the programming language, utilizing TensorFlow and PyTorch as the primary deep learning frameworks. Image processing was conducted using OpenCV and SimpleITK, while training was carried out on NVIDIA GPUs (Tesla T4) in Google Colab Pro.

### 1. Model Architecture:

In order to improve its capacity, we have modified the U-Net model with attention gates. on specified areas in the image. Key architectural components included

:

- Encoder: an ImageNet pre-trained ResNet-34 backbone for robust feature extraction.
- Attention Blocks: Integrated into the network to refine feature maps by emphasizing infection regions and irrelevant areas suppressed.
- Decoder: Comprised of transposed convolution layers for precise upsampling and reconstruction of segmentation maps.

### 2. Loss Functions:

Multiple loss functions were experimented with to optimize model performance:

- Dice Loss: For maximizing the overlap between predicted and true ground truth masks.
- Binary Cross-Entropy (BCE) Loss: The focus is on pixel-wise classification.

- Combined BCE-Dice Loss: A hybrid balance between pixel-level accuracy and region-level accuracy.

- Log Dice Loss: Dice Loss, a logarithmic variant, to counteract small segmentation regions.

- Surface Loss: With signed distance maps for boundary error costs.

- Weighted BCE-Dice Loss: We applied weights to address class imbalance between infected and non-infected regions.

3. **Training and Optimization:**

Training was conducted with the Adam optimizer with a learning rate of 1e-4. The selection was based on a grid search with different optimizers (e.g., SGD, RMSprop) as well as evaluating different configurations and learning rates to achieve the highest convergence and performance. To maximize convergence, the cosine annealing scheduler varied the learning rate dynamically. Training was conducted for 25 epochs.

Overfitting was avoided by applying early stopping, and a batch size of 16 slices was used. To ensure robustness, cross-validation techniques were implemented, and 20 percent of the dataset was reserved for training, validation, and testing.

4. **Evaluation Metrics:**

The model's performance was evaluated using the following metrics:

- **Dice Coefficient:** Measures overlap between predicted and ground truth masks.

- **Intersection over Union (IoU):** Measures the ratio of intersection to union of predicted and true masks.

- **Binary Accuracy:** Provides the proportion of correctly classified pixels.

- **Mean IoU:** Measures the average intersection over union across all classes.

- **Average Symmetric Surface Distance (ASSD):** Measures the average boundary distance between predictions and ground truth masks.

- **Hausdorff Distance:** Calculates the maximum distance between the boundaries of predicted and ground truth masks.

- **Classification Report:** Offers precision, recall, and F1-score for both background and mask classes, with all being close to perfect precision and recall for the model's segmentation tasks.

5. **Advanced Loss Function Integration:**

A novel generalized loss function combining weighted BCE-Dice loss and surface loss was implemented. This function allowed dynamic weighting through an alpha parameter, optimizing the balance between region-level and boundary-level accuracy.

6. **Callbacks and Checkpoints:**

Callbacks, including model checkpoints, early stopping, and learning rate schedulers were used for optimal model performance and storage efficiency.

**Results:**

1. **Results Without Augmentation Technique:**

The dataset without augmentation proved the model to be very effective for segmentation, yet it lacked the ability to handle data variation. Binary accuracy reached 99.72%, and the training loss consistently decreased over 50 epochs.

A Dice coefficient of 0.8502 and a mean IoU of 0.7445 were obtained in quantitative evaluation. Reasonably accurate boundary alignment was demonstrated with an average symmetric surface distance of 0.3907 and a hausdorff distance of 8.4853. However, without augmentation, training on the insufficiently augmented dataset resulted in lower accuracy and generalization to unseen data compared to the augmented set.

Results of the non-augmented model yielded an ROC curve (AUC = 0.91), exhibiting a strong but perhaps not as robust discriminatory capability compared to the augmented results. In Figure 4, we see that the IoU distribution was more varied and less consistent in terms of segmentation performance over the test dataset.

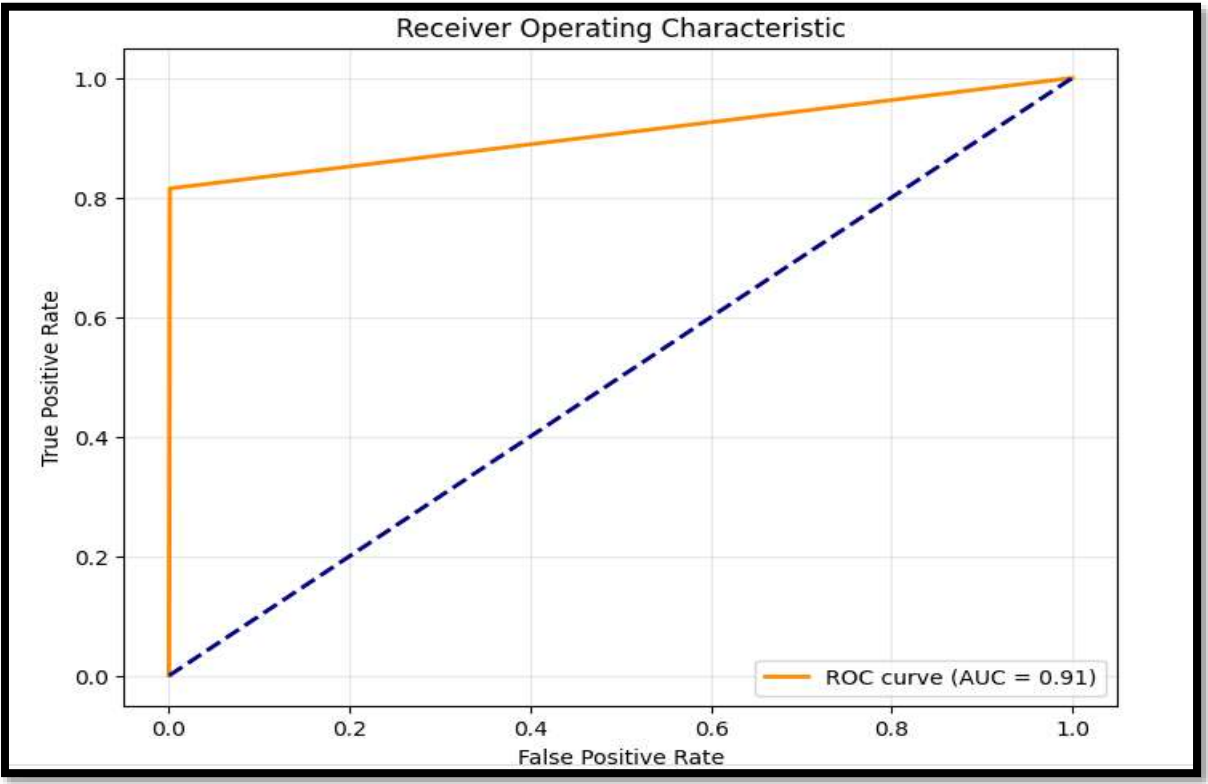

**Fig4.ROC Curve for Non-Augmented Model**

## 2. Results of Augmentation Technique:

Finally, it was shown that performance on the augmented dataset vastly outperformed the original dataset with respect to segmentation accuracy as well as boundary precision. As shown in Figure 5, training metrics consistently demonstrated reduced loss along with increased accuracy until 50 epochs; the final binary accuracy was 99.72%.

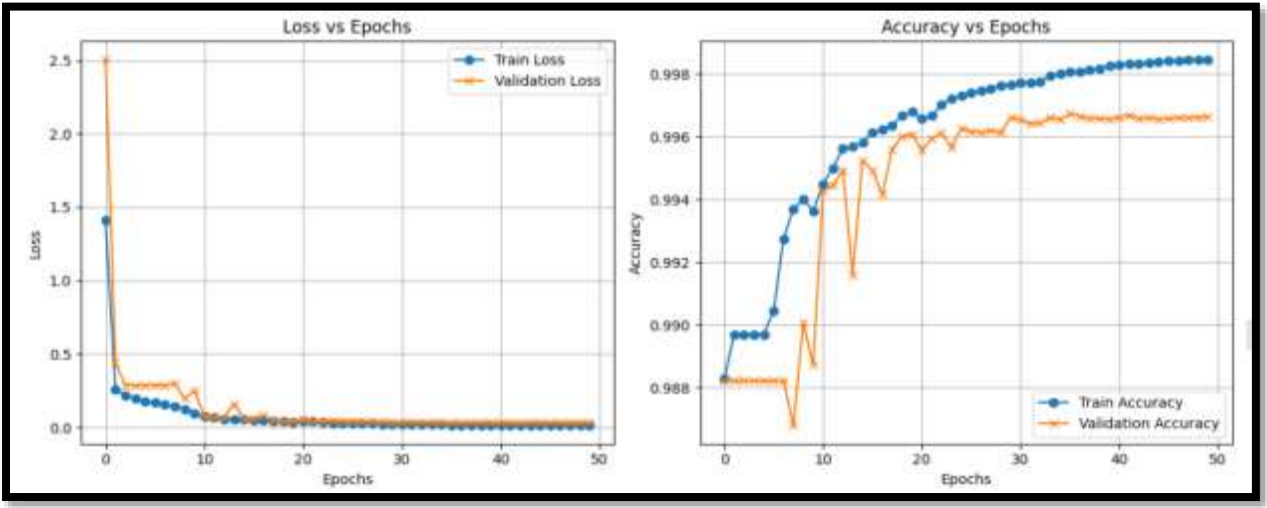

**Fig5. Training Metrics for Augmented Dataset: Loss and Accuracy Trends Across 50 Epochs**

Visually comparing original CT slices, ground truth masks, predicted masks, and post-processed masks, the model's segmentation results were explained. The high accuracy of the model was confirmed by quantitative evaluation, yielding a Dice coefficient of 0.8658 and mean IoU of 0.8316, as shown in Figure 6.

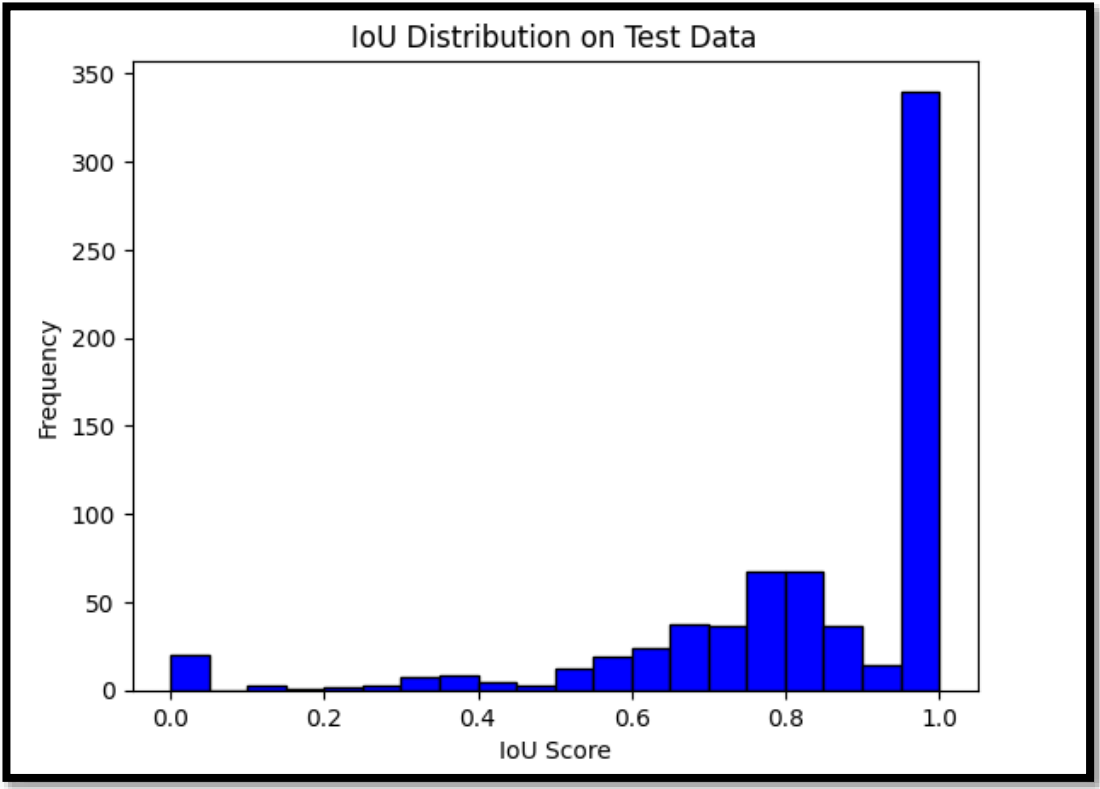

**Fig6. Frequency Distribution of IoU Values for Augmented Dataset**

An average symmetric surface distance (ASSD) of 0.3888 and a Hausdorff distance of 9.8995 were also included as additional metrics, confirming that the boundary prediction is very close to the ground truth boundary. The weighted precision, recall, and F1-score were 1.00 for the classification report as well. This is consistent with Figure 7, which shows the classification accuracy for both mask and background classes.

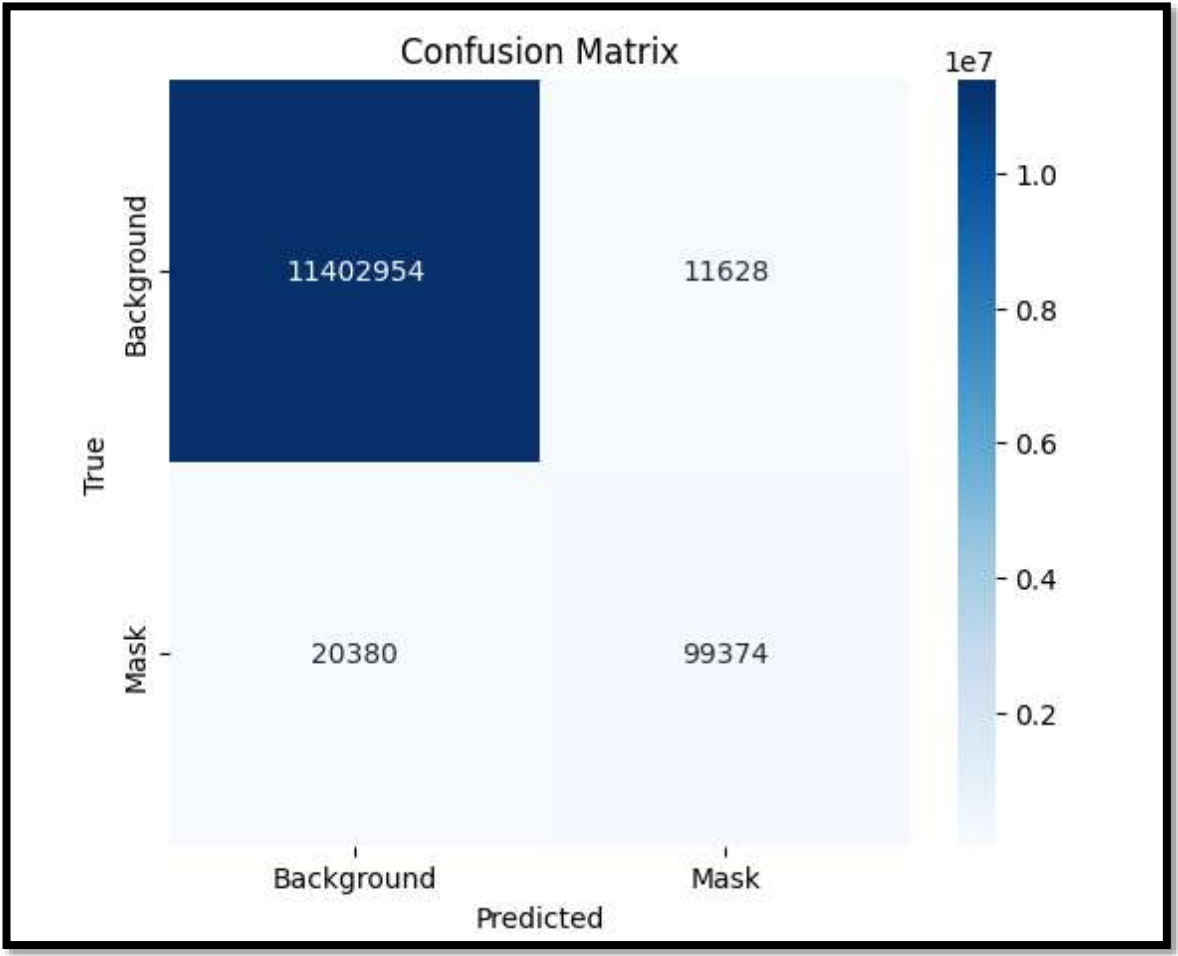

**fig7. Confusion Matrix for Mask and Background Classes**

The receiver operating characteristic (ROC) curve achieved an area under the curve (AUC) of 1.00,

reflecting the model's outstanding discriminatory capability, , as illustrated in Figure 8.

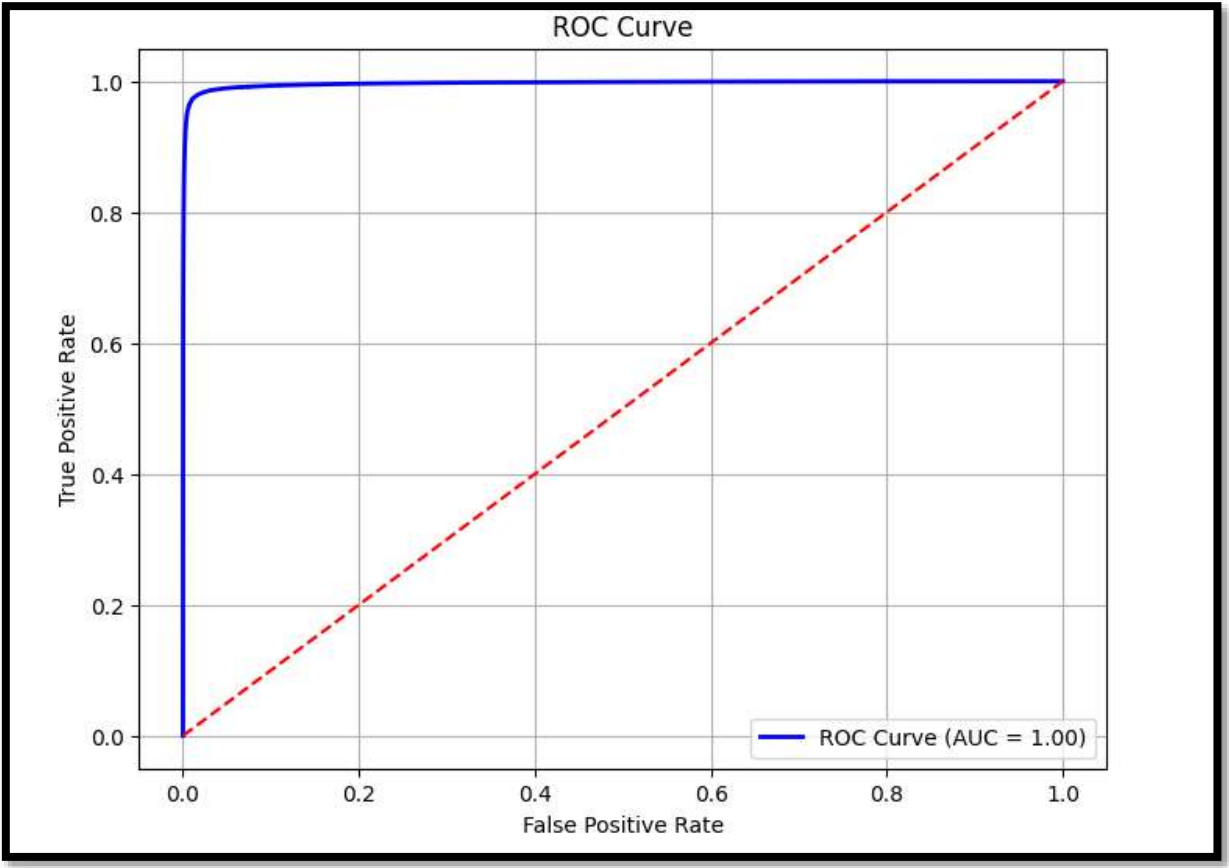

**Fig8. Receiver Operating Characteristic (ROC) Curve for Augmented Dataset with AUC = 1.00**

Comparisons between original CT slices, ground truth masks, and predicted masks (Figure 9A) suggest that the model can accurately delineate infection regions. As shown in Figure 9B, further improvements are possible with postprocessing.

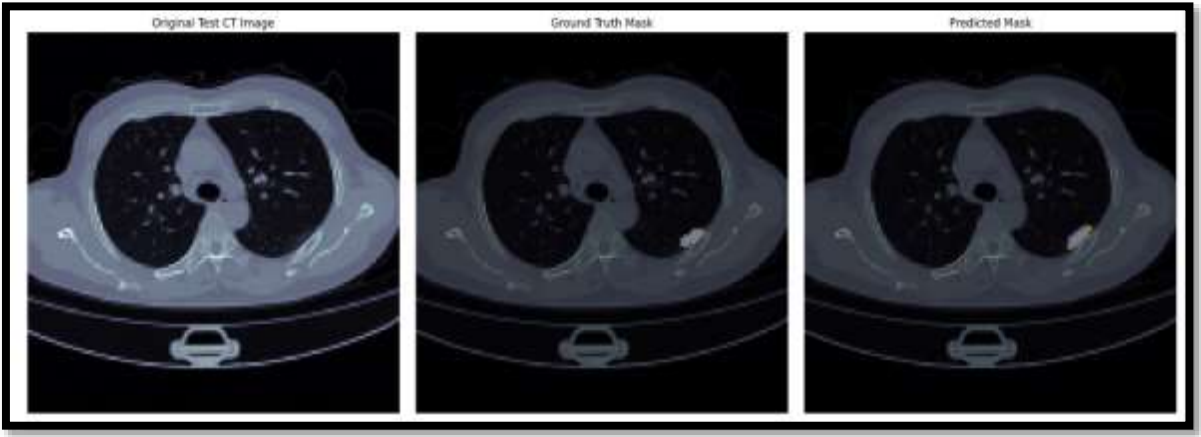

**Figure 9A: Comparison of Original CT Slices, Ground Truth Masks, and Predicted Masks**

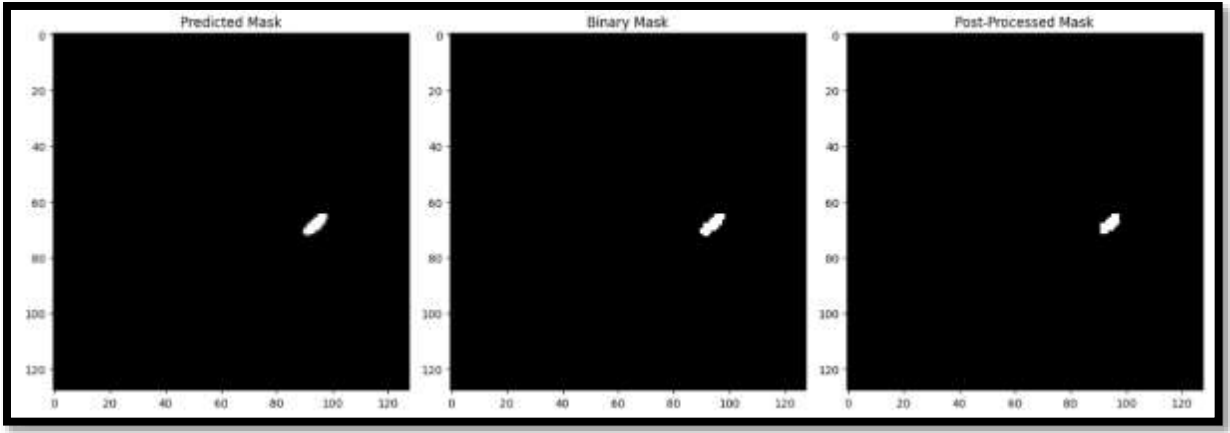

**Figure 9B: Post-Processed Masks Highlighting Refined Segmentation Results.**

### *Comparison of Results*

This difference between the results of non-augmented and augmented data indicates the effect of data augmentation on the accuracy and robustness of segmentation. It was demonstrated that the augmentation of data diversity resulted in an additional Dice coefficient improvement from 0.8502 to 0.8658 and a mean IoU improvement from 0.7445 to 0.8316. Compared to the results of the enhanced acquisition, similar ASSD, Hausdorff Distance, and improved boundary precision were obtained. The AUC of the ROC curve, initially 0.91, was significantly improved to 1.00 with an increased level of discrimination. Augmenting the dataset reduced overfitting, allowing the model to generalize more effectively to unseen data.

**Discussion:**

Data augmentation and advanced architectures for deep learning are shown to mitigate challenges in medical image segmentation in this study. Results showed that augmentation helped increase the accuracy of segmentation and enhance boundary precision, as reflected by metrics such as the Dice coefficient, ASSD, and Hausdorff Distance. This work would not have been possible without contributions from researchers like Jun Ma and colleagues, who created the COVID-19 CT Lung and Infection Segmentation Dataset. The annotations provided by their expert annotators offered a strong basis for training and evaluation. The proposed methodology outperformed baseline models in addressing variations in infection patterns and boundaries. Finally, attention mechanisms and post-processing were integrated further to improve the robustness and clinical applicability of the model.

***Comparison to Related Work***

The table below compares the results of this work to other studies that used the same dataset for COVID-19 lung region segmentation, along with references for further exploration:

| Study | Dice Coefficient | IoU | ASSD | Hausdorff Distance | Key Features | Reference |
|-------|------------------|-----|------|--------------------|--------------|-----------|
| Jun Ma et al. (2020) | 0.8345 | 0.7204 | 0.500 | 10.34 | Traditional U-Net architecture | (9) |
| Inf-Net (Deng-Ping Fan et al.) | 0.7800 | 0.6800 | 0.520 | 11.20 | Incorporation of edge-attention mechanisms | (10) |
| COVID-CT-Net (Jinyu Zhao et al.) | 0.8200 | 0.7400 | 0.450 | 9.50 | Ensemble learning and transfer learning | (11) |
| COVID-Rate (Nastaran Enshaei et al.) | 0.8500 | N/A | N/A | N/A | Automated segmentation framework with CT dataset | (12) |
| Automated 3D U-Net Segmentation (Dominik Müller et al.) | 0.8320 | 0.7280 | N/A | N/A | On-the-fly patch generation, 3D U-Net | (13) |
| Comparative Study (Sofie Tilborghs et al.) | 0.8100 | N/A | N/A | N/A | Multi-center evaluation with various algorithms | (14) |
| This Work (Augmented) | **0.8658** | **0.8316** | **0.3888** | **9.8995** | Attention U-Net, data augmentation, post-processing | Current study |
| This Work (Non-Augmented) | **0.8502** | **0.7445** | **0.3907** | **8.4853** | Attention U-Net | current study |

**Table.1 the results of this work to other studies that used the same dataset for COVID-19 lung region segmentation**

**Conclusion and Future Work:**

Particular CNN advances were demonstrated in this study to enable the automation of infected lung region segmentation in COVID-19 CT scans. Using data augmentation, attention mechanisms, and robust loss functions, the proposed methodology provided high accuracy and reliability. Quantitative metrics (Dice coefficient and IoU) and qualitative visualizations demonstrated that the model accurately delineated infection regions and adapted to data variations beyond the originally available data.

Segmentation quality was further improved through a post-processing pipeline that refined boundaries and removed noise, making the method suitable for clinical applications. The proposed approach was compared against existing studies and demonstrated advantages in both segmentation accuracy and generalization capability.

Future work can extend the data set to include more varied representative cases across demographics and pathologies to make the model robust. Furthermore, 3D segmentation techniques in volumetric CT data might yield richer spatial context for further study of infection. Computational efficiency optimization is essential for real-time deployment in clinical settings, e.g., through model pruning or quantization. Additionally, explainable AI techniques can be integrated to provide trust to clinicians through insights into the model's decision-making process. At last, working with interdisciplinary teams of radiologists and computer scientists as well as public health experts could help develop, for example, all-encompassing diagnostic tools for pandemic readiness.